\documentclass[pra,twocolumn,floatfix,superscriptaddress,longbibliography,nofootinbib]{revtex4-1}

\usepackage{tikz}
\usepackage[braket, qm]{qcircuit}
\usetikzlibrary{shapes,arrows,positioning}
\usepackage[ruled]{algorithm}
\usepackage{algpseudocode}
\usepackage{siunitx}
\usepackage{physics}
\usepackage{amssymb,amsmath,amstext, amsthm}
\usepackage{graphicx}
\usepackage{epstopdf}
\usepackage{color}
\usepackage{bm}
\usepackage{enumerate}
\usepackage[T1]{fontenc}
\usepackage{bbold}
\usepackage{bbm}
\usepackage{latexsym}
\usepackage[colorlinks=true,citecolor=blue,linkcolor=magenta]{hyperref}
\usepackage{braket}
\usepackage{multirow}

\newtheorem{theorem}{Theorem}
\newtheorem{lemma}{Lemma}

\newcommand{\fattheta}{\boldsymbol{\theta}}

\newcommand{\Ebb}{\mathbb{E}}

\newcommand{\GC}{\mathcal{G}}

\renewcommand{\geq}{\geqslant}
\renewcommand{\leq}{\leqslant}

\newcommand{\Var}{\textnormal{Var}}

\renewcommand{\vec}[1]{\boldsymbol{#1}}  

\newcommand{\tot}{\text{tot}}

\makeatletter
\newcommand{\justified}{%
  \rightskip=10pt $\,$%
  \leftskip=10pt }
\makeatother

\begin{document}
\title{Adaptive shot allocation for fast convergence in variational quantum algorithms}

\author{Andi Gu}
\thanks{The first two authors contributed equally to this work.}
\affiliation{Theoretical Division, Los Alamos National Laboratory, Los Alamos, NM 87545, USA}
\affiliation{Department of Physics, University of California, Berkeley, CA 94720, USA}

\author{Angus Lowe}
\thanks{The first two authors contributed equally to this work.}
\affiliation{Department of Combinatorics and Optimization and Institute for Quantum Computing, University of Waterloo, Waterloo, Ontario N2L 3G1, Canada}


\author{Pavel A. Dub}
\affiliation{Chemistry Division, Los Alamos National Laboratory, Los Alamos, New Mexico 87545, USA}

\author{Patrick J. Coles}
\affiliation{Theoretical Division, Los Alamos National Laboratory, Los Alamos, NM 87545, USA}

\author{Andrew Arrasmith}
\affiliation{Theoretical Division, Los Alamos National Laboratory, Los Alamos, NM 87545, USA}


\begin{abstract}
Variational Quantum Algorithms (VQAs) are a promising approach for practical applications like chemistry and materials science on near-term quantum computers as they typically reduce quantum resource requirements. However, in order to implement VQAs, an efficient classical optimization strategy is required. Here we present a new stochastic gradient descent method using an adaptive number of shots at each step, called the global Coupled Adaptive Number of Shots (gCANS) method, which improves on prior art in both the number of iterations as well as the number of shots required. These improvements reduce both the time and money required to run VQAs on current cloud platforms. We analytically prove that in a convex setting gCANS achieves geometric convergence to the optimum. Further, we numerically investigate the performance of gCANS on some chemical configuration problems. We also consider finding the ground state for an Ising model with different numbers of spins to examine the scaling of the method. We find that for these problems, gCANS compares favorably to all of the other optimizers we consider.
\end{abstract}
\maketitle

\section{Introduction}


Quantum computing may unlock previously intractable computations for a variety of applications in the physical sciences, industry, and beyond. However, the quantum computers that will be available in the near term are limited by having few qubits as well as hardware noise that limits the number of operations that can be performed before the information being manipulated degrades. Variational quantum algorithms (VQAs)~\cite{cerezo2020variationalreview,bharti2021noisy} are a promising approach to near-term quantum computing and have been proposed for many applications, such as electronic structure~\cite{peruzzo2014variational}, optimization~\cite{farhi2014quantum}, dynamical simulation~\cite{cirstoiu2020variational,commeau2020variational,endo2020variational,yuan2019theory}, linear systems~\cite{bravo2020variational,huang2019near,xu2019variational}, quantum compiling~\cite{khatri2019quantum,sharma2019noise}, quantum foundations~\cite{arrasmith2019variational}, and quantum sensing~\cite{beckey2020variational,cerezo2021sub,koczor2020variational}. VQAs encode the problem at hand in terms of a cost function computed from measurements of a quantum state and then vary that state to search for a solution. By only requiring a state preparation without a lengthy set of manipulations afterward, VQAs typically require many fewer qubits as well as much shorter run times than traditional quantum algorithms.


The reduced need for quantum resources allowed by VQAs comes at the cost of needing to classically optimize the control sequence, or ansatz, used to prepare the quantum state. This means that the efficiency of the method is largely determined by the computational expense of performing this optimization, which can often be non-trivial. Indeed, it has been shown that in general the classical optimization of a VQA can have numerous local minima and thus be NP-hard~\cite{bittel2021training}. Furthermore, for some cost functions and ansatzes, these VQA landscapes can exhibit exponentially flat landscapes known as barren plateaus~\cite{mcclean2018barren,cerezo2021cost,holmes2021barren,holmes2021connecting,sharma2020trainability,pesah2020absence,larocca2021diagnosing,wang2020noise,arrasmith2020effect,arrasmith2021equivalence,cerezo2020impact}. Even in the absence of such problems, the cost of performing the optimization may be prohibitive unless the optimizer is chosen with care.



Important considerations in determining the feasibility of a VQA include both the (wall-clock) time and money spent to find the solution. The theoretical run-time primarily depends on the number of different circuits run and the total number of shots.\footnote{The wall-clock time potentially includes the time spent performing the classical optimization step, latency in communicating with the server controlling the quantum device, the overhead from switching the device to a new circuit, and the time it takes to execute each shot (state preparation and measurement)~\cite{sung2020using}. As choices such as parallelism and performing the optimization on the server can significantly reduce the first two of these factors, the latter ones dominate.} Additionally, some platforms for accessing quantum devices on the cloud, such as Amazon's Braket service, charge both by the number of different circuits run and by the number of shots~\cite{winick_hamaker_hansen_2018}. In order to realize time efficiency and affordability for VQAs, one needs an optimizer that uses few iterations and few shots. Ideally, this optimizer would also require a minimum amount of hyperparameter tuning as using multiple runs to tune hyperparameters can greatly increase the overall time and cost of running the VQA.



Off-the-shelf classical optimizers (Adam, Nelder-Mead, Powell, etc.) have been employed and benchmarked in VQA applications~\cite{lavrijsen2020classical,kubler2020adaptive,larose2019variational}. Their often poor performance has led researchers to explore a new field, quantum-aware optimizers~\cite{kubler2020adaptive,arrasmith2020operator,stokes2020quantum,koczor2019quantum,nakanishi2020sequential,parrish2019jacobi,sweke2020stochastic,koczor2020quantum,van2021measurement}, which aim to tailor the optimizer to the idiosyncrasies of VQAs.
In the VQA setting, optimizers that use gradient information in theory offer improved convergence over those that do not~\cite{harrow2019low}. However, different gradient components can correspond to non-commuting observables, making gradient descent for VQAs more expensive than in the classical setting where a gradient can be measured all at once. In addition, shot noise (due to the variance of quantum observables) can result in a high shot cost during the optimization~\cite{wecker2015progress}. These concerns have led to the proposal of optimizers based on either small, fixed numbers of shots for each gradient component~\cite{sweke2020stochastic} or adaptive shot numbers~\cite{kubler2020adaptive,arrasmith2020operator}. 


In this work, we introduce a new optimization method that adaptively allocates shots for the measurement of each gradient component at each iteration. This optimizer, which we call the global-Coupled Adaptive Number of Shots (gCANS) method, uses a criterion for allocating shots that incorporates information about the overall scale of the shot cost for the iteration. In addition to providing shot frugality and hyperparameter robustness, the gCANS rule for shot allocation also leads to provable fast convergence. The upshot is that gCANS can save time and money relative to prior art, which our numerical comparison confirms.

In the following we begin with reviewing concepts and previous work on stochastic gradient descent for VQAs. We next introduce the gCANS algorithm and show that it achieves fast (geometric) convergence under some assumptions. Finally, we present numerical comparisons of the performance and resource requirements of performing the variational quantum eigensolver (VQE) on chemical and condensed matter systems using gCANS and other state-of-the-art methods.\footnote{We compare gCANS against iCANS~\cite{kubler2020adaptive}, Adam, and stochastic gradient descent with a geometrically increasing number of shots.} 



\section{Background}
A standard approach to optimization problems is gradient descent, characterized by an update rule that iteratively moves a candidate solution `downhill' in parameter space with respect to the cost function:
\begin{equation}
    \vb*{\theta}^{(t+1)} = \vb*{\theta}^{(t)} - \alpha \grad f(\vb*{\theta}^{(t)}) \label{eq:gd-update}
\end{equation}
Due to shot noise, we only have access to a noisy estimate of $f(\vb*{\theta})$. Under this setting, gradient descent becomes stochastic gradient descent (SGD). We review several key concepts for SGD below. First, we introduce the parameter shift rule we use to construct unbiased estimators of the gradients, then we introduce the random operator sampling we use to compute the expectation values in those estimators. Next we review the Lipshitz continuity of VQA gradients, and finally how this property has been used to define previous shot adaptive methods.

\subsection{Unbiased Estimators of the Gradient}\label{sec:grad-est}
For many VQAs in which the cost function is an expectation, it is possible to derive an analytical form for the gradient in terms of expectations of other observables, the so-called parameter-shift rules \cite{mitarai2018quantum,schuld2019evaluating}. We will discuss an instance of this class of VQAs: the case where the circuit ansatz is composed of single qubit rotations parameterized by $\vb*{\theta}^{(t)}$ and fixed entangling gates (e.g., CNOT rotations), and the cost function is $f(\vb*{\theta}) := \expval{U^\dagger(\vb*{\theta}) \hat{A} U(\vb*{\theta})}{0}$. In this case, the parameter shift rule gives
\begin{equation}\label{eq:analytic_derivative}
    \partial_i f(\vb*{\theta}) = \frac{f(\vb*{\theta}+\frac{\pi}{2} \vb*{\delta}_i)-f(\vb*{\theta}-\frac{\pi}{2} \vb*{\delta}_i)}{2}
\end{equation}
where $\vb*{\delta}_i$ is an indicator vector for the $i$th component of the parameter vector. Although we cannot calculate the exact expectations required for this gradient, we can make use of a random variable $\vb*{g}(\vb*{\theta}^{(t)})$ that is an unbiased estimator for the true gradient i.e., $\mathbb{E}[\vb*{g}(\vb*{\theta}^{(t)})]=\grad f(\vb*{\theta}^{(t)})$. For convenience, we define $X_i^{(t)}$ to be a random variable whose sample mean is the gradient estimator $g_i(\vb*{\theta}^{(t)})$, and $s_i^{(t)}$ to be the number of samples of $X_i^{(t)}$ used in the estimation process. For the class of VQAs discussed above $X_i^{(t)}=\frac{A_{i,+}^{(t)}-A_{i,-}^{(t)}}{2}$, where $A_{i,\pm}^{(t)}$ is a single-shot estimator of the expectation $f(\vb*{\theta} \pm \frac{\pi}{2} \vb*{\delta}_i)$. Although $A_{i,\pm}^{(t)}$ can be constructed in a number of ways, we use here weighted random operator sampling.

\subsection{Weighted Random Operator Sampling}
Since the expectation value of the $\hat{A}$ operators introduced above often cannot be efficiently measured directly, we decompose it as a sum of directly measurable Pauli operators: $\langle \hat{A} \rangle=\sum_k c_k \langle \hat{P}_k \rangle$. Given a budget of $s_{tot}$ shots to measure $\langle \hat{A} \rangle$, there are a number of ways we may distribute these shots between the Pauli operators. The simplest scheme will split the number of shots evenly across each operator (called uniform deterministic sampling) -- however, the variance of the resulting estimator is suboptimal, since it is desirable to invest more shots in estimating the highly weighted operators. Weighted deterministic sampling (WDS) assigns
\begin{equation}
    s_k=\left\lfloor s_{tot} \frac{\abs{c_k}}{\sum_i \abs{c_i}} \right\rfloor \label{eq:wds}
\end{equation}
When $\Var[\hat{P}_k]$ is the same for all $k$, this strategy is optimal \cite{Rubin_2018}, however, this also sets a floor on the number of shots needed to get an unbiased estimate for $\langle \hat{A} \rangle$. Weighted random sampling (WRS) circumvents this requirement by introducing randomness, enabling us to construct an unbiased estimator for $\langle \hat{A} \rangle$ with even just one shot. It does this by sampling $s_k$ from a multinomial distribution
\begin{equation}
    s_k \sim \text{Multi}\qty(s_{tot}, p_k=\frac{\abs{c_k}}{\sum_i \abs{c_i}})
\end{equation}
This procedure has been shown to outperform both uniform and weighted deterministic sampling \cite{arrasmith2020operator} for VQAs, hence we use WRS in each optimizer we study.

\subsection{Lipschitz-Continuous Gradients}

An appropriate choice of the learning rate, $\alpha$, is crucial for a well-behaved gradient descent algorithm. A learning rate that is too high will lead to updates that overshoot the minimum, whereas one that is too small will lead to overly cautious updates; both adversely affect the convergence rate. One can define reasonable bounds for a good learning rate given an upper bound on the slope of the cost function~$f$.

The upper bound of the gradient is formalized with the notion of a Lipschitz-continuous gradient. We call the gradient of $f$ Lipschitz-continuous if there is some Lipschitz constant $L \ge 0$ such that
\begin{equation}
    \norm{\grad{f}(\vb*{\theta}_a) - \grad{f}(\vb*{\theta}_b)} \le L \norm{\vb*{\theta}_a - \vb*{\theta}_b} \label{eq:lip-cont}
\end{equation}
for all $\vb*{x}_1,\vb*{x}_2 \in \text{dom}(f)$, where $\norm{\cdot}$ denotes an $\ell_2$ norm. If Eqn. \eqref{eq:lip-cont} holds and we are provided with access to the exact gradient, setting $\alpha \le \flatfrac{2}{L}$ is sufficient to guarantee convergence using the basic update rule in Eqn. \eqref{eq:gd-update}. For problems in the form discussed in Section~\ref{sec:grad-est}, we can find a convenient upper bound for $L$ by decomposing the observable of interest $\hat{A}$ into a sum of Pauli operators: $\hat{A}=\sum_k c_k \hat{P}_k$. Since $f$ is upper bounded in absolute value by $\sum_k \abs{c_k}$, we see that any partial derivative can be upper bounded with a constant  $\frac{\sum_k \abs{c_k}-(-\sum_k \abs{c_k})}{2} = \sum_k \abs{c_k}$. If our parameter vector is $d$-dimensional, we then have that
\begin{equation}
    L\le d \sum_k \abs{c_k} \,. \label{eqn:l-bound}
\end{equation}

For a more refined bound on the Lipschitz constant of cost functions arising in VQAs we refer the reader to~\cite{sweke2020stochastic}. Since the cost function in VQAs typically has a Lipschitz-continuous gradient, we can use Lemma~\ref{lem:fund_lem_2} in Appendix~\ref{sec:theorem_proof} to lower bound the improvement in the cost function (conditional on the gradient estimator):
\begin{equation}\label{eq:gain_lower_bound}
    f(\vb*{\theta}^{(t)})-f(\vb*{\theta}^{(t+1)}) \ge \alpha \grad f(\vb*{\theta}^{(t)})^\intercal \vb*{g}(\vb*{\theta}^{(t)}) - \frac{L \alpha^2}{2} \norm{\vb*{g}(\vb*{\theta}^{(t)})}^2\,.
\end{equation}
Using $\mathcal{G}$ to denote this lower bound on the gain, we are typically interested in its expectation:
\begin{equation}
    \mathbb{E}[\mathcal{G}] = \qty(\alpha - \frac{L \alpha^2}{2}) \norm{\grad f(\vb*{\theta}^{(t)})}^2 - \frac{L \alpha^2}{2} \sum_i \frac{\sigma_i^2}{s_i}\,. \label{eqn:exp-gain}
\end{equation}
where $\sigma_i$ is the standard deviation of $X_i$.
\subsection{Shot Adaptive Methods}
With the above formalism in hand, we now review existing optimization methods that use adaptive shot counts. The coupled adaptive batch size (CABS) algorithm was among the first to vary the number of measurements (batch size in machine learning) to maximize the expected improvement in the loss function~\cite{balles2017coupling}. The figure of merit is here $\frac{\mathbb{E}[\mathcal{G}]}{s}$, where $s$ is the total number of shots taken in the gradient estimation. They find that this can be maximized by taking
\begin{equation}
    s = \frac{2L \alpha}{2-L \alpha} \frac{\Tr(\Sigma)}{\norm{\grad f}^2}. \label{eqn:cabs}
\end{equation}
Since $\Sigma$ and $\grad f$ are not accessible, CABS uses an estimator $\hat{\Sigma}$ that replaces $\Sigma$ in Eqn. \eqref{eqn:cabs}, and furthermore approximates $\norm{\grad f}^2$ by $\flatfrac{f}{\alpha}$ (assuming $\underset{\theta}{\min} f(\theta)=0$).

The iCANS \cite{kubler2020adaptive} (individual coupled adaptive number of shots), has recently been introduced as a natural extension of CABS to VQAs. The key improvement in iCANS is that the number of shots per gradient component is allowed to vary individually, rather than using a uniform shot count as in CABS. Defining $\mathcal{G}_i$ as the gain (i.e., decrease in cost function) associated with updating the $i$th parameter $\theta_i$, iCANS is designed to maximize the shot efficiency
\begin{equation}
    \gamma_i := \frac{\mathbb{E}[\mathcal{G}_i]}{s_i}; \ i=1,\ldots,m \label{eqn:icans-fom}
\end{equation}
for each individual component of the gradient. This results in a shot count:
\begin{equation}
    s_i = \frac{2L \alpha}{2-L \alpha} \frac{\sigma_{i}^2}{g_i^2}.
\end{equation}
This optimization method has shown to be a significant improvement (achieving lower cost with fewer shots) over conventional optimizers such as ADAM, sequential optimization by function fitting (SOFF)~\cite{nakanishi2020sequential}, and simultaneous perturbation stochastic approximation (SPSA) \cite{kubler2020adaptive}. In light of these improvements, we work within the iCANS framework (allowing number of shots to vary for each component of the gradient) to propose a new optimizer, which we call gCANS: global coupled adaptive number of shots.

\section{gCANS Shot Allocation Rule}
We propose a new approach to SGD that, like iCANS, allows the number of shots per gradient component to vary. However, rather than allowing them to vary independently, we now optimize our expected gain globally over the entire gradient vector. That is, rather than taking an individual efficiency as in \eqref{eqn:icans-fom}, our figure of merit is now:
\begin{equation}
    \gamma = \frac{\mathbb{E}[\mathcal{G}]}{\sum_{k=1}^d s_k}
\end{equation}
Using the first order optimality condition $\grad_{\vb*{s}} \gamma=0$ (further details provided in Appendix~\ref{sec:derivation_gcans}), we obtain the rule:
\begin{equation}
    s_i = \frac{2L \alpha}{(2-L \alpha)} \frac{\sigma_i \sum_{k=1}^d \sigma_k}{\norm{\grad f(\vb*{\theta})}^2} \label{eqn:gcans-rule}
\end{equation}
This results from a \textit{global} metric for efficiency, hence we term this shot count prescription global coupled adaptive number of shots (gCANS). 

In practice the true standard deviations of the components and magnitude of the gradient will not be accessible, especially not before taking the step. We deal with this by using exponential moving averages to forecast the $\sigma_i$'s and $||\nabla f(\fattheta)||^2$. For the full details of how to carry out SGD with the gCANS rule, see Algorithm~\ref{alg:gCANS}.

\begin{figure}
\begin{algorithm}[H]
\begin{algorithmic}[1]
\Statex \textbf{Input:} Learning rate $\alpha$, starting point $\fattheta_0$, min number of shots per estimation $s_{\min}$, number of shots that can be used in total $N$, Lipschitz constant $L$, running average constant $\mu$
\State initialize: $\fattheta \gets \fattheta_0 $, $s_{\tot} \gets 0$,
$\vec{s} \gets (s_{\min} ,... ,s_{\min})^T$, $\vec{\chi}' \gets (0,...,0)^T$,
$\vec{\xi}' \gets (0,...,0)^T$, $k\gets 0$
\While{$s_{\tot} < N$}
    \State $\vec{g}, \vec{S} \gets iEvaluate(\fattheta, \vec{s})$
\State $s_{\tot} \gets s_{\tot} + 2 \sum_i s_i$
\State $\vec{\chi}'_\ell \gets \mu \vec{\chi}_\ell + (1-\mu) \vec{g}_\ell$
\State $\vec{\xi}'_\ell \gets \mu \vec{\xi}_\ell + (1-\mu) \vec{S}_\ell$
\State $\vec{\xi}_\ell \gets \vec{\xi}'_\ell/(1-\mu^{k+1})$
\State $\vec{\chi}_\ell \gets \vec{\chi}'_\ell/(1-\mu^{k+1})$
\For{$ i \in [1,...,d]$}
        \State $\fattheta_i \gets \fattheta_i - \alpha \vec{g}_i$
    \State $s_i \gets \left\lceil\frac{2L\alpha}{2-L\alpha} \frac{\xi_i \sum_j \xi_j}{||\vec{\chi}||^2}\right\rceil$
\EndFor
\State $k\gets k + 1$
\EndWhile
\end{algorithmic}
\caption{\justified{Stochastic gradient descent with gCANS. The function {$iEvaluate(\fattheta, \vec{s})$} evaluates the gradient at $\fattheta$ using $s_i$ shots for measuring both of the expectation values needed to compute the $i$-th derivative via the parameter shift rule in~\eqref{eq:analytic_derivative}. This function returns the estimated gradient vector $\vec{g}$ as well as the vector $\vec{S}$ whose components are the variances of the estimates of the partial derivatives.}\label{alg:gCANS}}
\end{algorithm}
\end{figure}

\section{Convergence Analysis}\label{sec:convergence}
The gCANS update rule guarantees fast convergence in expectation to the optimal value of sufficiently smooth cost functions. Here, sufficiently smooth means the function is both strongly convex and has Lipschitz-continuous gradients, properties which are explained in more detail in Appendix~\ref{sec:theorem_proof}. Given our cost function satisfies these properties, the adaptively chosen shot sizes for each component of the gradient estimator ensure that the optimality gap $f(\vb*{\theta}^{(t+1)})-f^*$ is proportional to $O(2^{-t})$ after $t$ updates, which is referred to as \textit{geometric convergence}. In other words, only $O(\log(1/\epsilon))$ iterations are required to achieve optimality gap at most $\epsilon$. 

In the classical ML setting it is known that both decreasing the learning rate and increasing the number of samples used to estimate the gradient leads to convergence to the optimum, in expectation~\cite{bottou2018optimization}. In the context of VQAs, Sweke et al.~\cite{sweke2020stochastic} point out that SGD with a constant learning rate leads to fast convergence given that the variance of the estimator decays with the magnitude of the gradient (which would require increasing the number of shots as the optimization progresses). We make use of this fact and show that our update rule leads to sufficiently small variances for the estimator, keeping in mind that in VQAs the sample sizes can vary across different components of the gradient. This demonstrates that, unlike previous analysis in the classical ML setting, it is not necessary to increase the sample sizes geometrically over the different components of the gradient estimator to ensure the variances are sufficiently small. Rather, one may use a criteria for the different shot sizes which is informed by the cost function and estimators.

Before presenting the theorem, we state the following assumptions about the cost function $f$, learning rate $\alpha$ and estimators $\vb*{g}(\vb*{\theta}^{(t)})$ used in our gCANS updates:
\begin{enumerate}
    \item $\mathbb{E}[g_i(\vec{\theta}^{(t)})]= \partial_i f(\vec{\theta}^{(t)})\ \forall i \in [d], t\in \mathbb{N}$.
    \item $\Var[g_i(\vec{\theta}^{(t)})] = \frac{\Var[X_i^{(t)}]}{s_i^{(t)}} \ \forall t \in \mathbb{N}$.
    \item $f$ is $\mu$-strongly convex.
    \item $f$ has $L$-Lipschitz continuous gradient.
    \item $\alpha$ is a constant learning rate satisfying $0 < \alpha < \min\{1/L,2/\mu\}$.
    \item It holds that
    \begin{align*}
        s_i^{(t)} = \frac{2L\alpha}{2-L\alpha} \frac{\sigma_i^{(t)}\left(\sum_{j=1}^d \sigma_j^{(t)}\right)}{\norm{\nabla f (\vec{\theta}^{(t)})}^2} \ \forall i \in [d], t\in \mathbb{N}
    \end{align*}
    where $\sigma_i^{(t)} := \sqrt{\Var[X_i^{(t)}]}$.
\end{enumerate}
The final assumption represents an idealized version of the gCANS update rule, since we cannot in general know the gradient magnitude and estimator variances. We will refer to these assumptions in the proof of the theorem, which we defer to Appendix~\ref{sec:theorem_proof}. We can now state our theorem concerning fast convergence to the optimum.
\begin{theorem}[gCANS geometric convergence]\label{thm:gcans_convergence}
Under the assumptions given above, SGD updates achieve geometric convergence to the optimal value of the cost function. In other words,
\begin{align}
     \mathbb{E}[f(\vb*{\theta}^{(t)})] - f^*  = O\left(\gamma^t\right)
\end{align}
for some $0<\gamma<1$.
\end{theorem}

To summarize the implications of this theorem, under the aforementioned assumptions gCANS is guaranteed to converge quickly, approaching the optimal cost value exponentially quickly in the number of iterations. In realistic VQA applications we expect these assumptions to hold with the exception that VQA landscapes will be non-convex. However, if the optimizer settles into a convex region of the landscape we would then expect this fast convergence to be realized.

\section{Numerical Benchmarks}\label{sec:numerics}

\begin{figure}[t]
    \centering
    \begin{align*}
    \Qcircuit @C=0.7em @R=1em {
	 	 & \gate{R_Y} & \gate{R_Z} & \ctrl{1} & \qw & \qw & \gate{R_Y} & \gate{R_Z} & \qw & \qw \\
	 	 & \gate{R_Y} & \gate{R_Z} & \targ & \ctrl{1} & \qw & \gate{R_Y} & \gate{R_Z} & \qw & \qw\\
	 	 & \gate{R_Y} & \gate{R_Z} & \qw & \targ & \ctrl{1} & \gate{R_Y} & \gate{R_Z} & \qw & \qw\\
	 	 &  &  & & &  & & & &  \\
	 	&  &  & & & \vdots & & & &  \\
	 	 & \gate{R_Y} & \gate{R_Z} & \qw & \qw & \targ & \gate{R_Y} & \gate{R_Z} & \qw & \qw \gategroup{1}{2}{6}{6}{.7em}{--}
	 } \\
	 k=1,\ldots,\mathcal{D}-1 \qquad \qquad \qquad \qquad
\end{align*}
    \caption{Circuit ansatz of depth $\mathcal{D}$, termed the `hardware-efficient SU(2) 2-local circuit'. There are a total of $2\mathcal{D}$ parameters, one for each $Y$ and $Z$ single-qubit rotation.}
    \label{fig:ansatz}
\end{figure}
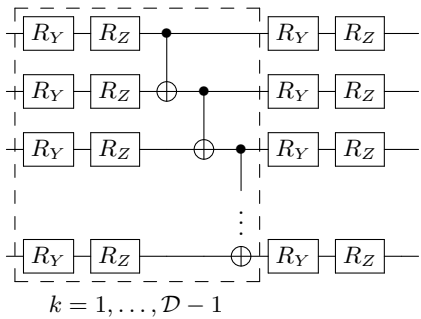

To compare gCANS with existing methods, we apply it to a variational quantum eigensolver (VQE) for three physical systems: He$_2^+$ (5 qubits), NH$_3$ (12 qubits), and a transverse field Ising model (4 to 10 qubits). For all numerical experiments, we use the circuit ansatz shown in Fig. \ref{fig:ansatz}, with a varying depth for each of the three problems. 

We compare four optimizers: gCANS, iCANS, ADAM, and stotchastic gradient descent with dynamic sampling (hereafter called SGD-DS). By dynamic sampling, we mean the number of shots expended per iteration is set to increase by a fixed geometric schedule: $s=\lfloor s_0 r^k \rfloor$ where $s_0$ is the initial number of shots for each component of the gradient. Furthermore, we set the learning rate to be $\alpha=\frac{0.5}{L}$ for iCANS, ADAM, and SGD-DS for VQE on He$_2^+$ and NH$_3$. We find gCANS operates better at a slightly higher learning rate $\alpha=\frac{1}{L}$ for these two problems (see Section~\ref{sec:disc}). For the Ising model scaling analysis, since $L$ grows (approximately linearly) with system size, we set $\alpha=\frac{0.5}{L}$ for both iCANS and gCANS (since we find both to be relatively insensitive to learning rate for this case). The hyperparameters for ADAM are set to the recommended values $\beta_1=0.9$, $\beta_2=0.99$, and $\epsilon=10^{-8}$, for iCANS we have set $\mu=0.99$ and $b=10^{-6}$, and finally for gCANS we set $\mu=0.99$ as well. Any remaining free hyperparameters were chosen empirically on a per-problem basis to yield the best performance for each optimizer.

We use Qiskit~\cite{Qiskit} with the PySCF backend~\cite{PySCF} to calculate the Hamiltonian operators for each of the physical problems we consider, and TensorFlow~\cite{tensorflow2015-whitepaper} to efficiently vectorize our simulations.

\begin{figure}[t!]
    \includegraphics[width=0.48\textwidth]{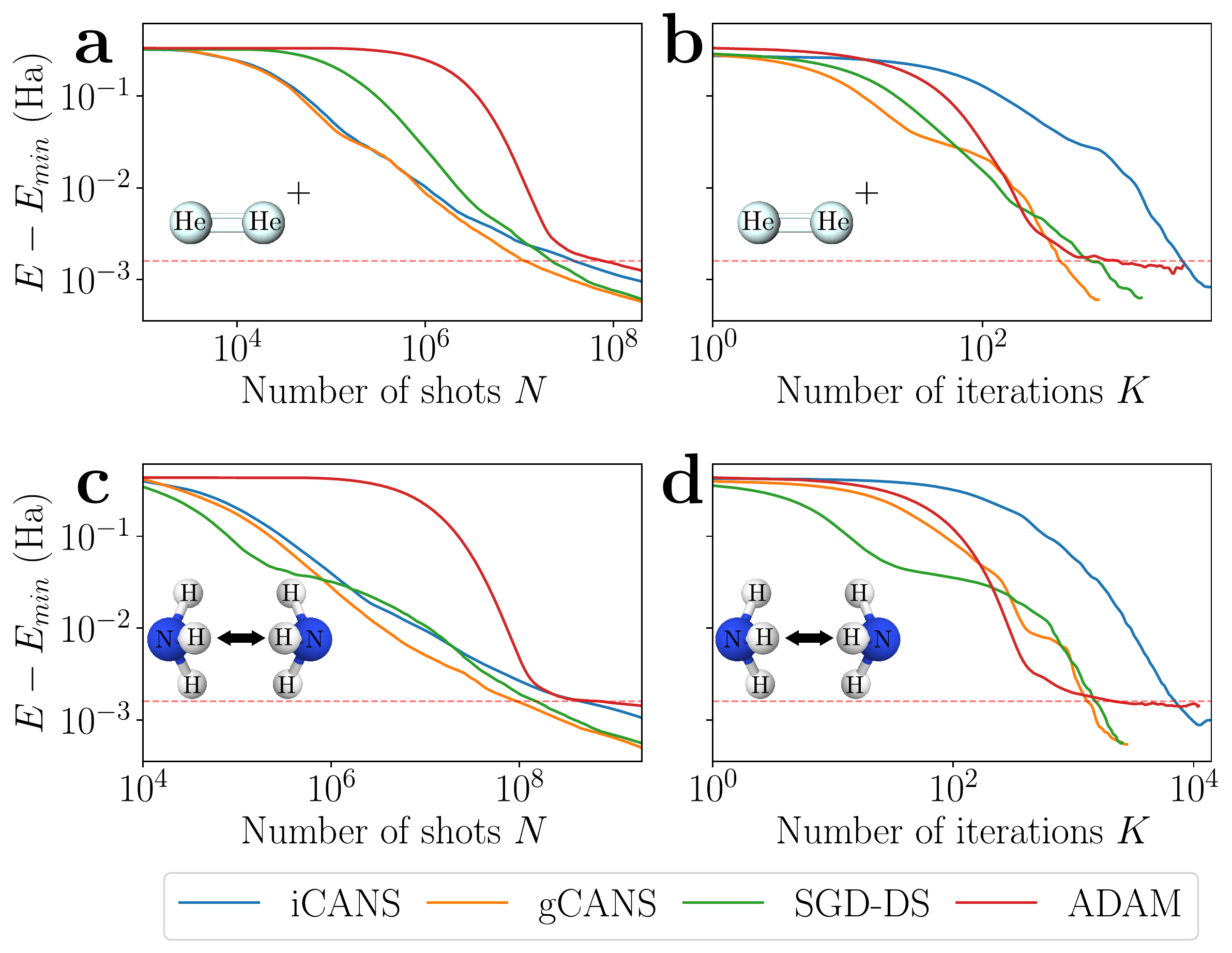}
    \caption{Cost as a function of number of shots expended over 10 random starts, at the equilibrium position of He$_2^+$ and NH$_3$ in panels \textbf{a} and \textbf{c}, respectively. The cost as a function of number of iterations is shown for both He$_2^+$ and NH$_3$ in panels \textbf{b} and \textbf{d}, respectively. Chemical accuracy ($\num{1.6e-3}\text{Ha}$) is marked by a horizontal dashed red line. We emphasize the comparable performance of gCANS and SGD-DS is realized only when the common ratio for the latter is very carefully tuned (see Appendix~\ref{sec:hyp-sensitivity}).}\label{fig:training-curve}
\end{figure}

\begin{figure}[!ht]
    \centering
    \includegraphics[width=0.4\textwidth]{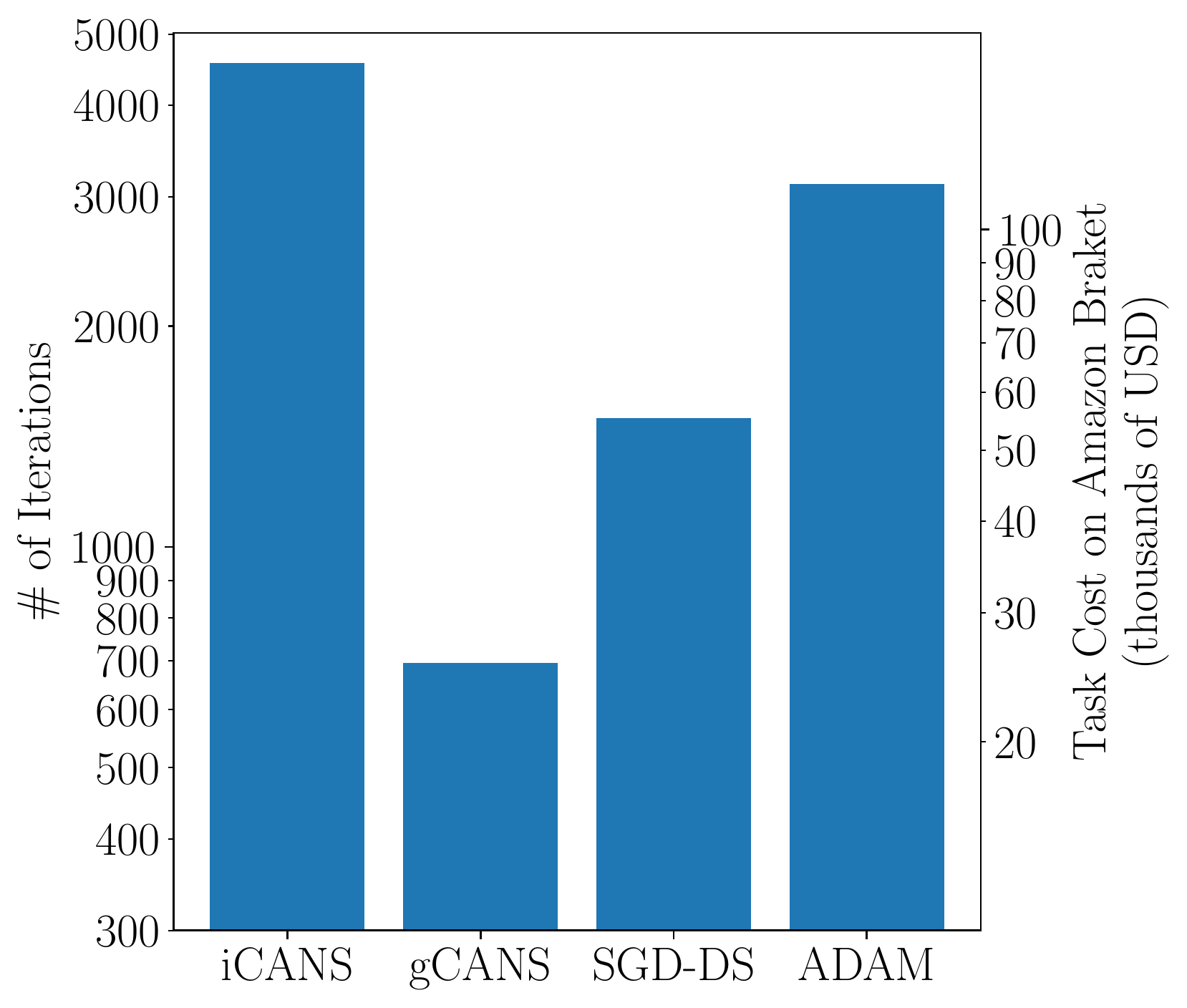}
    \caption{Number of iterations taken over the full course of VQE shown in Figure ~\ref{fig:training-curve} for He$_2^+$ at the equilibrium distance. The corresponding task cost (not including costs due to number of shots expended) for running these problems on Amazon Braket is shown on the right. Note that this differs from the data in Table~\ref{tab:iters-costs} as the optimizers were not stopped when they reached chemical accuracy -- they were stopped when they reached their budget of $\num{2e8}$ shots.}
    \label{fig:iters-cost}
\end{figure}

\begin{table*}
    \centering
    \begin{tabular}{|c| c | c | c | c| c|}
        \hline
        Problem & Optimizer & Iterations ($K$) & Shots ($S$) & Cost (thousands of USD) & Estimated Time (hours)\\ \hline \hline
        \multirow{4}{2em}{He$_2^+$} & iCANS & 3015 & \num{4.6e7} & 127 & 12.86\\
        & \textbf{gCANS} & $\boldsymbol{353}$ & $\boldsymbol{1.4 \times 10^7}$ & $\boldsymbol{18}$ &$\boldsymbol{1.98}$ \\
        & SGD-DS & 853 & \num{3.5e7} & 44 & 4.86 \\
        & ADAM & 1450 & \num{8.7e7} & 84 & 9.79\\ \hline
        \multirow{4}{2em}{NH$_3$} & iCANS & 8301 & \num{5.1e8} & 5968 & 564.44 \\
        & \textbf{gCANS} & $\boldsymbol{1243}$ & $\boldsymbol{9.8 \times 10^7}$ & $\boldsymbol{901}$ & $\boldsymbol{85.72}$\\
        & SGD-DS & 1395 & \num{1.8e8} & 1036 & 100.10 \\
        & ADAM & 2740 & \num{5.5e8} & 2104 & 207.51 \\ \hline
    \end{tabular}
    \caption{We list the number of iterations and number of shots (averaged over 10 random starts) required to reach chemical accuracy for the four optimizers we compare for the equilibrium configurations for He$_2^+$ and NH$_3$. We also include the corresponding hypothetical cost, in USD, if these optimizers were to be used for VQE on Amazon Braket. The costs are computed with $C=0.3P K + 0.00035 S$ dollars, where $K$ is the number of iterations, $P$ is the number of Pauli terms in the expansion of $\hat{H}$, and $S$ is the total number of shots used  \cite{winick_hamaker_hansen_2018}. We estimate the wall clock time by assuming shots are taken with a frequency of $5$ kHz (the sampling rate of~\cite{google2019supremacy}) and that it takes $0.1$ seconds to change the circuit being run (following~\cite{sung2020using}). This gives a time estimate $T=0.1P K + 0.0002 S$ seconds. We neglect latency or time spent on classical update steps. Neither the cost or the time estimates accounts for the burden of the hyperparameter tuning, which would likely be substantial for performing SGD-DS.}
    \label{tab:iters-costs}
\end{table*}

\begin{figure*}
     \includegraphics[width=\textwidth]{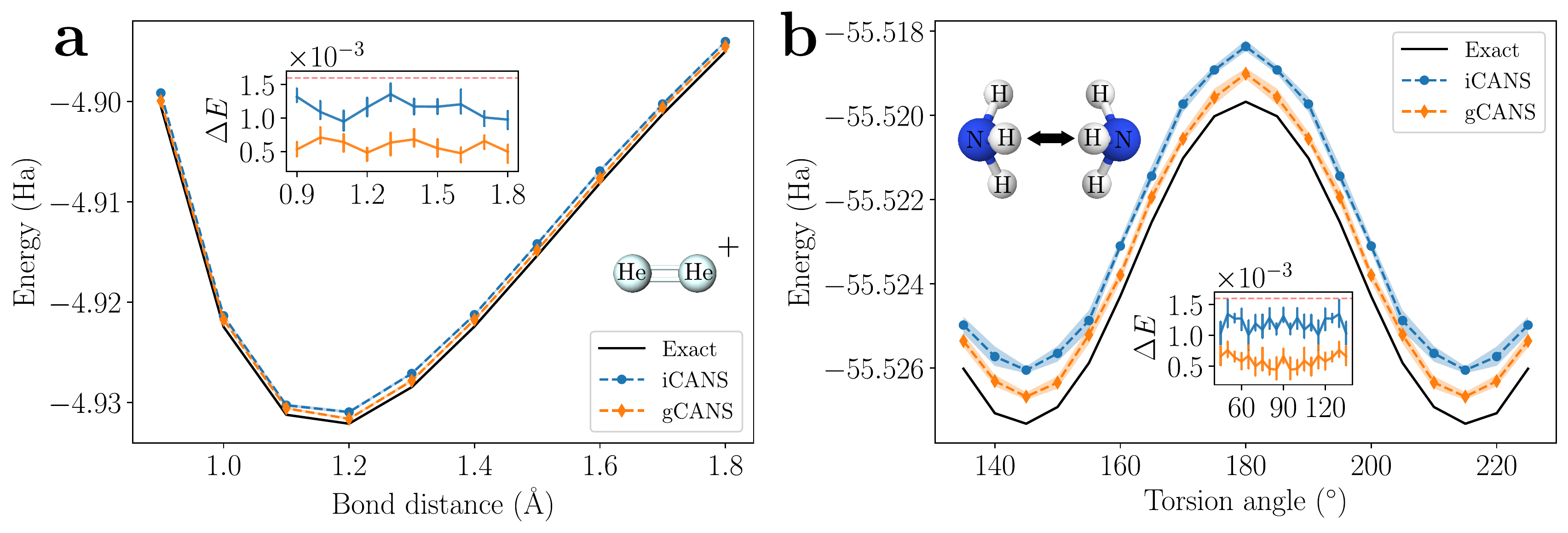}
    \caption{Potential energy curves as a function of bond distance for He$_2^+$ (panel \textbf{a}) and torsion angle for NH$_3$ (panel \textbf{b}), as calculated using iCANS and gCANS with a fixed depth circuit ansatz. A 95\% confidence interval is shown in the shaded regions. The total number of shots used for each point on the He$_2^+$ and NH$_3$ curves were $2 \times 10^8$ and $2 \times 10^9$, respectively. The insets show the difference from the exact diagonalization energy, with the dashed red line indicating chemical accuracy $\epsilon=\num{1.6e-3}\text{Ha}$.}
    \label{fig:bonddistance}
\end{figure*}

\subsection{Dihelium Cation and Ammonia Umbrella Inversion}

For He$_2^+$, we use the 6-31G basis set to encode the molecular Hamiltonian into 8 qubits via parity mapping. We further reduce the qubit count to 5 by applying the well-known two qubit reduction and other $\mathbb{Z}_2$ symmetries~\cite{bravyi2017tapering}. Our circuit ansatz has depth $\mathcal{D}=6$. For SGD with dynamic sampling, we set the initial number of shots per gradient component to $s_0=500$ and $r=1.0025$. For ADAM, we set the number of shots per gradient component to be 2500. We apply a similar qubit tapering procedure for ammonia using the STO-3G basis set, and use a circuit ansatz with depth $\mathcal{D}=10$. For SGD with dynamic sampling, we set $s_0=1500$ an $r=1.001$. For ADAM, we set the number of shots per gradient component to be 5000.

\subsection{Scaling Comparison}
\begin{figure*}
    \centering
    \includegraphics[width=0.85\textwidth]{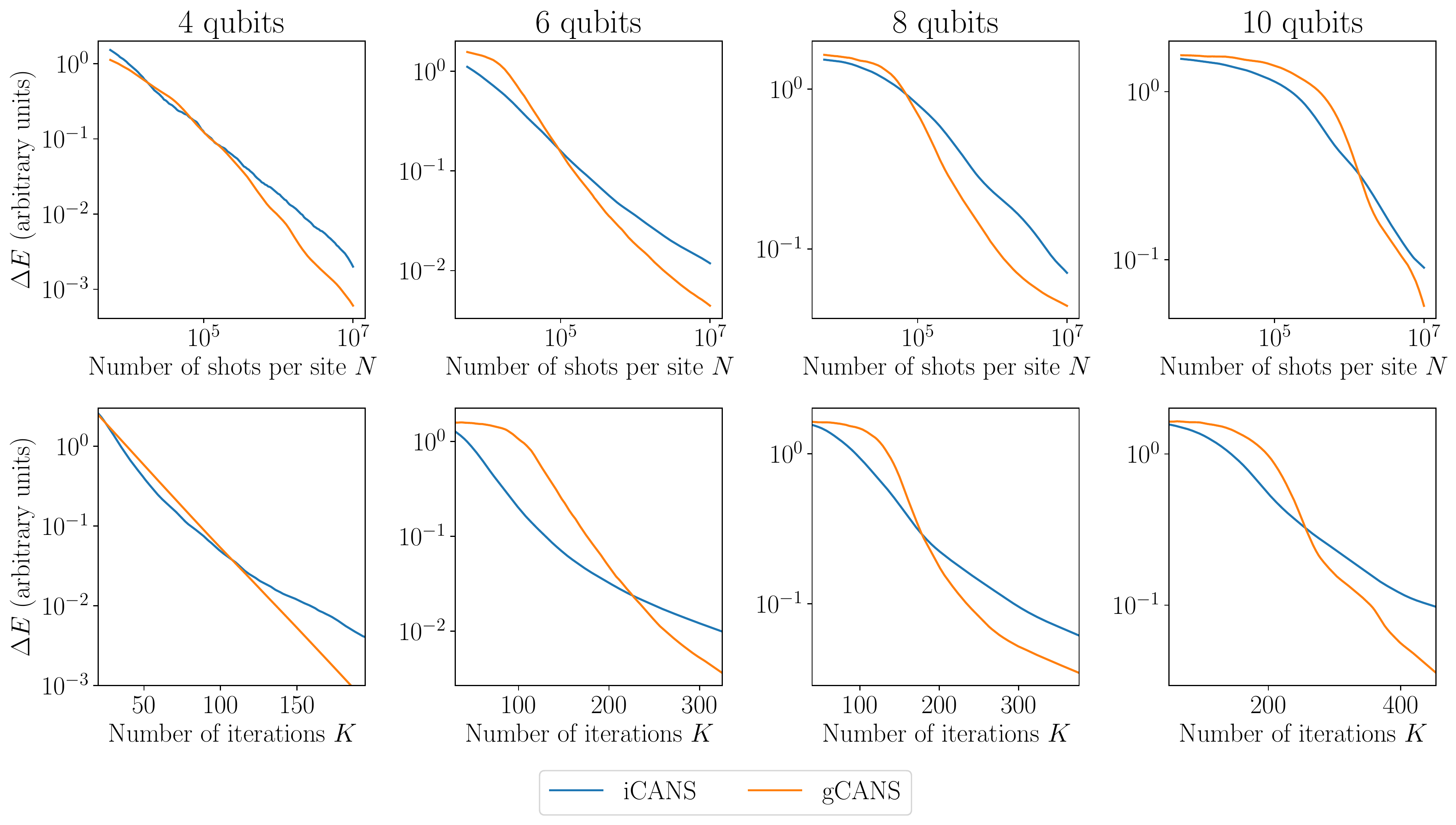}
    \caption{Performance of iCANS and gCANS on VQE for the transverse field Ising model for 4, 6, 8, and 10 qubits.}
    \label{fig:scaling}
\end{figure*}

Since gCANS is most similar in form to iCANS, we do a head-to-head scaling comparison of these two optimizers by applying both to VQE for a transverse field Ising model:
\begin{equation}
    H = \sum_{\langle i, j \rangle} Z_i Z_i + g \sum_i X_i
\end{equation}
where $\langle i, j \rangle$ denotes nearest neighbour sites $i$ and $j$. We set $g=1.5$ for each of our experiments. Finally, we use a circuit depth of $\mathcal{D}=6$. The results for $n=4,6,8$ and $10$ qubits are shown in Fig. \ref{fig:scaling}.

\section{Numerical Results}\label{sec:results}

Here we summarize the results of our numerical benchmarks. In the following discussion, we highlight the resource requirements in terms of the number of iterations ($K$) and number of shots ($S$), which both impact the run time and monetary cost of the algorithm. We also discuss the robustness in terms of the sensitivity to hyperparameters, as the overhead associated with hyperparameter tuning can also be viewed as impacting the run time and monetary cost.

For the quantum chemistry tasks we considered, we found that for both the dihelium ion and ammonia inversion cases that gCANS achieves chemical accuracy with fewer shots and iterations than the other optimizers (see Fig.~\ref{fig:training-curve}). When the hyperparameters are well tuned, SGD-DS is the most competitve, typically achieving comparable performance to gCANS. However, we show in Appendix~\ref{sec:hyp-sensitivity} that SGD-DS is highly sensitive to the choice of hyperparameters while gCANS is not, and thus SGD-DS would likely require additional shots and iterations to tune. In comparison, Adam is less sensitive than SGD-DS to the choice of hyperparameters, but requires more shots and iterations to reach chemical accuracy. Finally, iCANS is again more robust the choice of hyperparameters and uses a number of shots comparable to or a little better than Adam but needs by far the most iterations to reach chemical accuracy.

Figure~\ref{fig:iters-cost} shows the number of iterations used in the whole optimization for the dihelium at the equilibrium bond length as well as the corresponding task price that would be charged on the Amazon Braket system for running this number of tasks. One can clearly see the financial benefit of using gCANS over the other considered optimizers in this figure. Table~\ref{tab:iters-costs} summarizes the total cost and estimated run times involved in reaching chemical accuracy for both chemistry benchmarks, and gCANS is the top performer in all categories in this table.


Figure~\ref{fig:bonddistance} provides an additional comparison between gCANS and iCANS for a fixed shot budget.  Here we see that for all bond distances for the dihelium ion, and for all torsion angles for ammonia, gCANS gets closer to the true ground state energy than iCANS does. We also see that gCANS achieves energies well below the chemical accuracy threshold.

Figure~\ref{fig:scaling} analyzes the performance as the problem size scales up, for the transverse Ising model ground state problem. Naturally, the cost landscape becomes more difficult to navigate as the problem size increases, and consquently we observe some degradation in the performance (for both iCANS and gCANS) with increasing problem size. However, regardless of the problem size considered, we observe the same general trend: gCANS outperforms iCANS both in terms of shot resources and iteration resources.

\section{Discussion}\label{sec:disc}


In this work we have proposed a new shot frugal optimization strategy for variational quantum algorithms (VQAs), which is well-motivated by fast convergence guarantees. We call this method global Coupled Adaptive Number of Shots  (gCANS). gCANS, like its predecessor iCANS, provides a dynamic allocation of measurements (shots) taken on a quantum computer for the determination of each gradient component, but differs in how it does so. The rule used by gCANS to determine the optimal number of shots in each update considers the shot cost for the entire gradient rather than just the cost of just a single component, which leads to our fast convergence result (Theorem~\ref{thm:gcans_convergence}).


To understand the characteristics and benefits of the gCANS method, we compared its performance to three other gradient-based optimizers: ADAM, stochastic gradient descent with dynamic sampling, and iCANS. In addition, given its similarities with the iCANS method, we made a direct head-to-head comparison of gCANS and iCANS for a larger variety of problems, including a study on the scaling behavior of both for a transverse field Ising model. 


 First, the ADAM optimizer was found to be competitive with iCANS and gCANS up to the point where it could not make further progress. This result is consistent with previous findings \cite{kubler2020adaptive}, and represents the thematic problem for optimization in VQAs: using a large number of shots early in the optimization is wasteful, since we can make substantial progress even with a noisy gradient when we are far from the minimum. On the other hand, being too frugal later on prevents progress since we require a precise estimate of the gradient as we approach the minimum. This necessitates some degree of dynamism for any shot-frugal optimizer: the remaining three optimizers we compare all have dynamic shot counts.

Stochastic gradient descent with dynamic sampling exhibits the qualitative behavior we expect from a shot frugal optimizer: it starts using very few shots, and steeply increases number of shots towards the end of the optimization process. Indeed, we find that after tuning the hyperparameters of stochastic gradient descent with dynamic sampling, we can achieve comparable performance to gCANS, as shown in Fig. \ref{fig:training-curve}. However, this comes at the cost of significant investment in finding the optimal hyperparameters $s_0$ and $r$. The performance is extremely sensitive to the choice of $r$ (see Appendix~\ref{sec:hyp-sensitivity}). For settings of $r$ that are barely suboptimal, SGD-DS can perform much worse than even ADAM. Moreover, we have not found any generally applicable rule for selecting $r$ -- that is, it must be chosen \textit{ad hoc} for each problem. 

The iCANS optimizer is most directly comparable to gCANS, since they are operating under the same framework, the key difference being the gCANS shot count rule in~\eqref{eqn:gcans-rule}. iCANS has already been shown to outperform various popular optimizers~\cite{kubler2020adaptive}, in agreement with our findings here. Beyond its shot frugality, iCANS carries the additional advantage of requiring little to no hyperparameter tuning -- the small amount of tuning we did to find a good learning rate was found to only marginally affect performance. However, since a well-tuned SGD-DS optimizer outperforms iCANS, we suspect in general that the iCANS shot allocation rule is suboptimal. 


We contend gCANS ought to supersede iCANS as the shot-frugal optimizer of choice for VQAs, as it inherits the strengths of iCANS while addressing its weaknesses. We summarize these desirable characteristics:
\begin{enumerate}[i)]
    \item gCANS consistently outperforms each of the optimizers we test, achieving chemical accuracy with fewer shots and fewer circuit compilations. This translates to faster and cheaper (see Table \ref{tab:iters-costs}) experiments, bringing us closer to a practical implementation of VQE on near-term quantum computers.
    \item Similar to iCANS, gCANS is extremely robust to changes in its hyperparameters (see Appendix~\ref{sec:hyp-sensitivity}), unlike optimizers such as SGD-DS. This robustness reduces the resources required to identify the appropriate settings of these hyperparameters.
    \item Unlike iCANS, which does not lead to an optimal shot allocation in any setting (however idealized), gCANS has attractive convergence rate guarantees (see Theorem \ref{thm:gcans_convergence}).
\end{enumerate}
Finally, our scalibility analysis (Fig. \ref{fig:scaling}) indicates that the advantages listed above will hold as we increase the number of qubits. We therefore believe that gCANS will play an important role in the quest towards near-term quantum advantage.


\section{ACKNOWLEDGEMENTS}

 Research presented in this article was supported by the Laboratory Directed Research and Development (LDRD) program of Los Alamos National Laboratory (LANL) under project number 20200056DR. AG and AL acknowledge support from the U.S. Department of Energy (DOE) through a quantum computing program sponsored by the LANL Information Science \& Technology Institute.  PJC also acknowledges support from the LANL ASC Beyond Moore's Law project. AA was also initially supported by the LDRD program of LANL under project number 20190065DR. 

\bibliographystyle{unsrt}
\bibliography{gcans,quantum}

\onecolumngrid
\appendix

\section{Derivation of gCANS update rule}\label{sec:derivation_gcans}
Define $\GC$ to be the lower bound on the total gain~\cite{Balles2017} from an update step as in the right-hand side of~\eqref{eq:gain_lower_bound}
\begin{align}
    \GC := \left(\alpha - \frac{L\alpha^2}{2}\right)\grad f(\vec{\theta})^\intercal \vec{g} - \frac{L\alpha^2}{2}\norm{\vec{g}}^2
\end{align}
and consider the task of maximizing this quantity divided by the number of shots taken. We can solve for the optimal number of shots to estimate each component of the gradient
\begin{align}
    \underset{\vec{s}}{\text{argmax}} \ \frac{\mathbb{E}[\GC]}{\sum_{k=1}^ds_k}
\end{align}
using the first-order optimality condition
\begin{align}
    \grad_{\vec{s}} \left(\frac{\mathbb{E}[\GC]}{\sum_{k=1}^ds_k}\right) = 0.
\end{align}
This gives us the following set of coupled equations
\begin{align}
    \frac{1}{\sum_{k=1}^ds_k}&\left[\left(\alpha - \frac{L\alpha^2}{2}\right)\norm{\grad f(\vec{\theta})}^2 - \frac{L\alpha^2}{2}\sum_{k=1}^d\frac{\sigma_k^2}{s_k}\right] \nonumber \\
    &= \frac{L\alpha^2}{2}\frac{\sigma_i^2}{s_i^2} \quad \forall\ i \in \{1,\dots,d\}\label{eqn:coupled_condn}
\end{align}
from which one obtains
\begin{align}
    s_j = \frac{\sigma_j}{\sigma_k} s_k \quad \forall\ j,k \in \{1,\dots,d\}.
\end{align}
Consider a fixed index $i$. Substituting the above relationship for every $s_j \neq s_i$ in (\ref{eqn:coupled_condn}) yields
\begin{align}
    \frac{1}{\sum_{k=1}^d\sigma_k}&\left[\left(\alpha - \frac{L\alpha^2}{2}\right)\norm{\grad f(\vec{\theta})}^2 - \frac{L\alpha^2}{2}\frac{\sigma_i}{s_i}\sum_{k=1}^d\sigma_k\right] \nonumber\\
    &= \frac{L\alpha^2}{2}\frac{\sigma_i}{s_i}\\
    \implies &\left(\alpha - \frac{L\alpha^2}{2}\right)\norm{\grad f(\vec{\theta})}^2 = L\alpha^2\frac{\sigma_i}{s_i}\sum_{k=1}^d\sigma_k.
\end{align}
Solving for $s_i$, we arrive at
\begin{align}
    s_i = \frac{2L\alpha \ \sigma_i\sum_{k=1}^d\sigma_k}{(2-L\alpha)\norm{\grad f(\vec{\theta})}^2}
\end{align}
for which the denominator is strictly greater than zero (except at exact optima) so long as the learning rate is $\alpha < 2/L$.
\section{Proof of Theorem \ref{thm:gcans_convergence}}\label{sec:theorem_proof}
Let us first recall that a function $f: \mathbb{R}^d\to \mathbb{R}$ is $\mu$-strongly convex for a constant $\mu>0$ if
\begin{align}
    f(\vec{y})\geq f(\vec{x}) + \nabla f(\vec{x})^\intercal (\vec{y}-\vec{x}) + \frac{1}{2}\mu\norm{\vec{y}-\vec{x}}^2
\end{align}
for every $\vec{x},\vec{y}\in \mathbb{R}^d$. Before stating the proof of Theorem \ref{thm:gcans_convergence}, it will also be helpful to review the following intermediate results. We refer the reader to Section 4 of \cite{bottou2018optimization} for further discussion on the role of strong convexity as well as proofs for Lemmas \ref{lem:pl_inequality} and \ref{lem:fund_lem_2}. The first lemma is based on a condition known as the Polyak-\L{}ojasiewicz (PL) Inequality~\cite{karimi2020linear}, introduced in \cite{Polyak1963}.
\begin{lemma}[PL Inequality]\label{lem:pl_inequality}
For a $\mu$-strongly convex function $f : \mathbb{R}^d \to \mathbb{R}$ with optimum $f^*$, it holds that
\begin{align}
    2\mu(f(\vec{x}) - f^*) \leq \norm{\grad f(\vec{x})}_2^2 \quad \forall \vec{x} \in \mathbb{R}^d
\end{align}
\end{lemma}
\begin{lemma}[Lemma 4.2 in \cite{bottou2018optimization}]\label{lem:fund_lem_2}
The iterates of the stochastic gradient descent algorithm for an objective function $f$ with $L$-Lipschitz continuous gradient satisfy
\begin{align}
    \Ebb[f(\vec{\theta}^{(k+1)})] - f(\vec{\theta}^{(k)}) \leq -\alpha \grad f(\vec{\theta}^{(k)})^\intercal \Ebb[\vec{g}(\vec{\theta}^{(k)})] + \frac{1}{2}\alpha^2 L \Ebb\left[\norm{\vec{g}(\vec{\theta}^{(k)})}^2\right]
\end{align}
where the expectation is taken over the random variables in the $k^\text{th}$ iteration.
\end{lemma}
We can now begin the proof of the theorem, referring to Assumptions 1-6 in Section~\ref{sec:convergence} as needed. Under Assumptions 1 and 4 and using Lemma \ref{lem:fund_lem_2}, the iterates satisfy
\begin{align}
    \Ebb[f(\vec{\theta}^{(k+1)})] - f(\vec{\theta}^{(k)}) \leq -\alpha \norm{\grad f(\vec{\theta}^{(k)})}^2 + \frac{1}{2}\alpha^2 L \Ebb\left[\norm{\vec{g}(\vec{\theta}^{(k)})}^2\right]
\end{align}
where expectations are over the random variables in the $k^\text{th}$ iteration. Also, it holds that
\begin{align}
    \Ebb\qty[\norm{\vec{g}(\vec{\theta}^{(k)})}^2] = \sum_{i=1}^d \Var[g_i(\vec{\theta}^{(k)})] + \norm{\grad f(\vec{\theta}^{(k)})}^2
\end{align}
so that, combining with Assumption 2 we get
\begin{align}
    \Ebb\qty[f(\vec{\theta}^{(k+1)})] - f(\vec{\theta}^{(k)}) \leq -\left(\alpha-\frac{1}{2}\alpha^2L\right) \norm{\grad f(\vec{\theta}^{(k)})}^2 + \frac{1}{2}\alpha^2 L\sum_{i=1}^d \frac{(\sigma_i^{(k)})^2}{s_i^{(k)}}.
\end{align}
Now, noting that our shot allocation rule (Assumption 6) implies
\begin{align}
    \frac{\qty(\sigma_i^{(k)})^2}{s_i^{(k)}} = \frac{(2-L\alpha)\norm{\grad f(\vec{\theta}^{(k)})}^2\  \sigma_i^{(k)}}{2L\alpha\left(\sum_{j=1}^d \sigma_j ^{(k)} \right)}
\end{align}
the inequality may be written as
\begin{align}
    \Ebb\qty[f(\vec{\theta}^{(k+1)})] - f(\vec{\theta}^{(k)}) &\leq -\left(\alpha-\frac{1}{2}\alpha^2L\right) \norm{\grad f(\vec{\theta}^{(k)})}^2 + \frac{1}{2}\left(\alpha - \frac{1}{2}\alpha^2L\right)\norm{\grad f (\vec{\theta}^{(k)})}^2\nonumber\\
    &= -\frac{1}{2}\left(\alpha - \frac{1}{2}\alpha^2L\right)\norm{\grad f(\vec{\theta}^{(k)})}^2\nonumber\\
    &\leq -\frac{1}{4}\alpha\norm{\grad f (\theta^{(k)})}^2
\end{align}
where the last line made use of Assumption 5. Next, using the $\mu$-strong convexity of $f$ (Assumption 3) and Lemma \ref{lem:pl_inequality}, we have
\begin{align}
    \Ebb\qty[f(\vec{\theta}^{(k+1)})] - f(\vec{\theta}^{(k)}) &\leq -\frac{1}{2}\alpha\mu\left(f(\vec{\theta}^{(k)}) - f^*\right).
\end{align}
Finally, adding $f(\vec{\theta}^{(k)}) - f^*$ to both sides and taking the total expectation over all iterations (and still using $\Ebb$ to denote this), one obtains
\begin{align}
    \Ebb\qty[f(\vec{\theta}^{(k+1)})] - f^* \leq \left(1-\frac{1}{2}\alpha\mu\right)\left(\Ebb[f(\vec{\theta}^{(k)})] - f^*\right)
\end{align}
which, by Assumption 5 is less than or equal to $\gamma \left(\Ebb\qty[f(\vec{\theta}^{(k)})] - f^* \right)$ for some $\gamma < 1$ so, by induction,
\begin{align}
    \Ebb\qty[f(\vec{\theta}^{(k+1)})] - f^* &\leq R \gamma^k
\end{align}
for $R := \underset{\vec{\theta}}{\max}\ f(\vec{\theta}) - f^*$.
\section{Hyperparameter Sensitivity}\label{sec:hyp-sensitivity}
In Section \ref{sec:numerics}, we present results for optimizers that have been carefully tuned to the each specific problem. In practice, it is not always possible to do this meta-optimization. It is desirable, therefore, to use an optimizer that is robust to changes in their hyperparameters.
\begin{figure*}[t]
    \includegraphics[width=.9\textwidth]{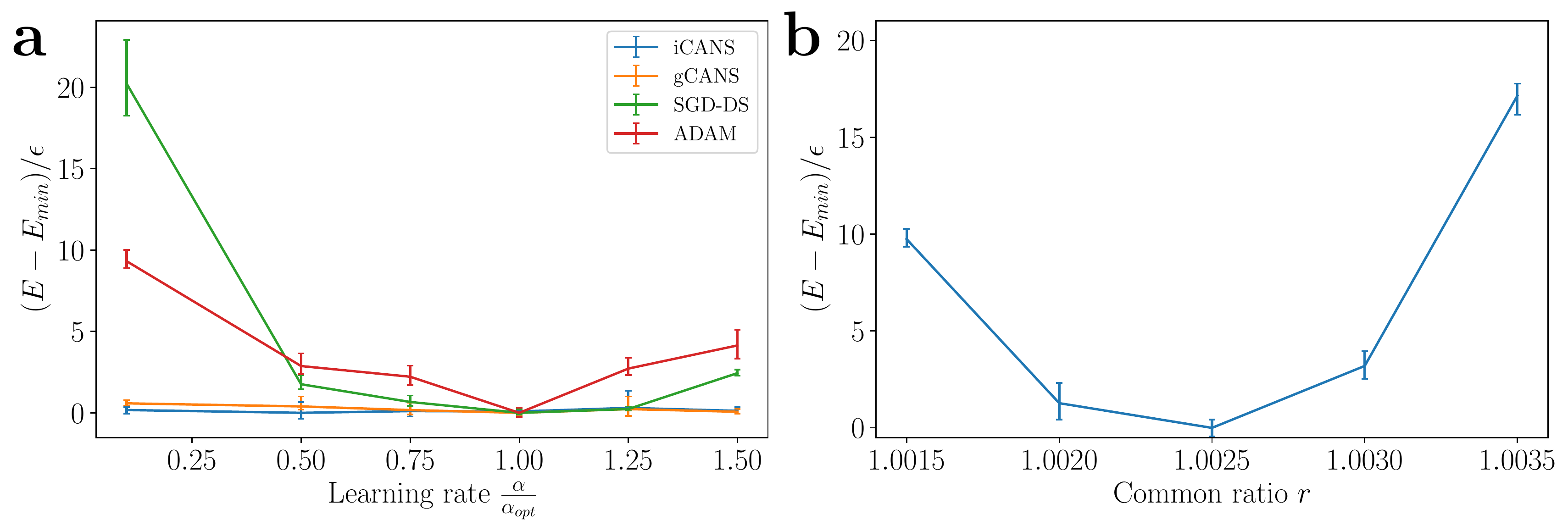}
    \caption{The learning rate sensitivity of various optimizers on the He$_2^+$ VQE problem (panel \textbf{a}) and the sensitivity of SGD-DS to its common ratio (panel \textbf{b}). The $y$-axis is the difference (normalized by $\epsilon=\num{1.6e-3}\text{Ha}$) between the energy after $2 \times 10^8$ shots and the the minimum energy found by each optimizer. Our results in Fig. \ref{fig:training-curve} use the optimal hyperparameters $\alpha_{opt}=\frac{0.5}{L}$ for iCANS, ADAM, and SGD-DS, $\alpha_{opt}=\frac{1}{L}$ for gCANS and $r=1.0025$ for SGD-DS.}
    \label{fig:hyperparam-sensitivity}
\end{figure*}

We find that iCANS and gCANS, both being adaptive optimizers, are very robust to changes in their learning rate (see Fig.~\ref{fig:hyperparam-sensitivity}). Notably, gCANS typically benefits from a higher learning rate compared to iCANS -- we believe this is due to iCANS being overly conservative in its shot allocation, primarily due to the prescription to clip $s_i$ to a maximum value. Although this clipping is necessary later in the optimization, in the early phases of noisy, fast descent, this clipping is too restrictive. 

In contrast to adaptive optimizers such as iCANS and gCANS, ADAM and and SGD-DS are quite sensitive to their learning rate (when we hold all other hyperparameters fixed). Given the strong relationship between optimal shot count \eqref{eqn:gcans-rule} and learning rate, this high degree of sensitivity is to be expected in non-adaptive optimizers. We also find SGD-DS is very sensitive to the common ratio $r$ in its shot schedule (see Fig.~\ref{fig:hyperparam-sensitivity}), and furthermore find no general prescription for setting this ratio: for the He$_2^+$ VQE problem, an optimal $r=1.0025$ was used, whereas for NH$_3$ the optimal was $r=1.001$. Therefore, we emphasize the relatively strong performance of SGD-DS in Fig.~\ref{fig:training-curve} is not robust. 
\end{document}